\documentclass[aip,jcp,amsmath,amssymb,reprint,superscriptaddress]{revtex4-2}
\usepackage[dvipdfmx]{graphicx}
\usepackage{epsf}
\usepackage{bm}
\usepackage{physics}
\usepackage{lipsum}
\usepackage[normalem]{ulem}
\usepackage{txfonts}
\usepackage{color}

\def\ls{\lesssim}

\graphicspath{{./}{./figures}}

\begin{document}

\title{Elucidation of the Correlation between Molecular Conformation
and Shear Viscosity of Polymer Melts under Steady-State Shear Flow}

\author{Yuhi Sakamaki}
\affiliation{Division of Chemical Engineering, Department of Materials Engineering Science, Graduate School of Engineering Science, The University of Osaka, Toyonaka, Osaka 560-8531, Japan}

\author{Shota Goto}
\affiliation{Division of Chemical Engineering, Department of Materials Engineering Science, Graduate School of Engineering Science, The University of Osaka, Toyonaka, Osaka 560-8531, Japan}

\author{Kang Kim}
\email{kk@cheng.es.osaka-u.ac.jp}
\affiliation{Division of Chemical Engineering, Department of Materials Engineering Science, Graduate School of Engineering Science, The University of Osaka, Toyonaka, Osaka 560-8531, Japan}

\author{Nobuyuki Matubayasi}
\email{nobuyuki@cheng.es.osaka-u.ac.jp}
\affiliation{Division of Chemical Engineering, Department of Materials Engineering Science, Graduate School of Engineering Science, The University of Osaka, Toyonaka, Osaka 560-8531, Japan}

\begin{abstract}
The rheological behavior of polymer melts is strongly influenced by
parameters such as chain length, chain stiffness, and architecture. 
In particular, shear
thinning, characterized by a power-law decrease in shear viscosity with
increasing shear rate, has been widely investigated through 
molecular dynamics simulations. 
A central question is the connection
between molecular conformation under steady flow and the resulting
shear-thinning response. 
In this study, we employed coarse-grained
molecular dynamics simulations of linear and ring polymers with varying chain
stiffness to examine this relationship, with chain conformations
quantified by the gyration tensor.
We identified a strong correlation between the velocity-gradient direction component
of the gyration tensor and shear viscosity, which exhibits a clear scaling
relationship. 
This indicates that chain extension along the velocity-gradient direction governs
the effective frictional force. 
Notably, this behavior emerges as a general feature, independent of
chain architecture and chain stiffness. 
In addition, shear viscosity was found to correlate with the component of the
gyration tensor corresponding to the direction that is not directly influenced by advective effects of shear flow. 
Because advection is absent in the direction, polymer chains can be
regarded as diffusing freely, and the extent of this diffusion appears
to be controlled by the shear viscosity.
\end{abstract}

\maketitle

\section{INTRODUCTION}

Polymer melts exhibit shear thinning, characterized by a decrease in
shear viscosity $\eta$ with increasing shear rate $\dot{\gamma}$, described by
$\eta \sim\dot{\gamma}^{-\alpha}$, where $\alpha$ is the shear-thinning
exponent.~\cite{doi1986Theory} 
This non-linear rheological behavior has primarily been investigated for
linear polymers of varying chain
lengths using non-equilibrium molecular dynamics (NEMD)
simulations.~\cite{kroger1993Rheology, aust1999Structure,
aoyagi2000Molecular, kroger2000Rheological, xu2014Shear}
Under shear flow, polymer chains are expected to elongate along the flow
direction while contracting along the velocity-gradient direction. 
Consequently, understanding shear thinning from the perspective of
molecular conformation is essential. 
Notably, although the shear-thinning power-law exponent depends on
polymer architecture, simulation studies have examined only whether 
relationships exist between shear viscosity and structural descriptors that
characterize molecular conformation.

Xu, Chen, and An reported NEMD simulation results of 
linear, ring, star, and H-shaped polymer melts, demonstrating the universal
relationship between zero-shear viscosity and mean-squared gyration of
radius, expressed as $\eta_0 \propto \langle R_g^2\rangle$.~\cite{xu2015Simulation}
They further examined the relationship between polymer structure and
shear viscosity under 
steady-state shear flow.~\cite{xu2017Probing}
Specifically, a simple relationship, $\eta \propto \langle
G_{yy}\rangle^{3/2}$, was demonstrated in both the linear and nonlinear
rheological regimes.
Here, $\langle G_{yy}\rangle$ represents the $y$-component of 
the gyration tensor, which is calculated by expressing each monomer bead's
position as a vector relative to the 
chain's center of mass and averaging the corresponding tensor products.
Among the resulting components, $\langle G_{xx}\rangle$,
$\langle G_{yy}\rangle$, and $\langle G_{zz}\rangle$ represent the squared ellipsoidal radii along the
$x$, $y$, and $z$ axes when the polymer chain is approximated as an ellipsoid.
Note that the $x$-axis corresponds to the flow direction, and the
$y$-axis corresponds to the velocity-gradient direction.

Nikoubashman and Howard investigated the effect of chain stiffness on
the rheological properties and molecular conformations of linear polymer
melts.~\cite{nikoubashman2017Equilibrium} 
They found that increasing chain stiffness leads to higher zero-shear
viscosity $\eta_0$, 
which is attributed to enhanced intermolecular steric interactions. 
In contrast, at high shear rates, the rheological response becomes
largely independent of chain stiffness. 
Analysis of polymer conformations under shear revealed that, at high
shear rates, chains exhibit a pronounced tendency to elongate and align
along the flow direction.
They also identified scaling relationships,  $\langle
G_{xx}\rangle \propto \mathrm{Wi}^{-0.45}$ and $\langle
G_{yy}\rangle \propto \mathrm{Wi}^{-0.3}$, where the Weissenberg number
is defined as $\mathrm{Wi}= \dot{\gamma} \tau_\mathrm{c}$ with the
time scale $\tau_\mathrm{c}$ determined from the mean-squared displacement
of the monomers, $g_1(t)$, such that $g_1(\tau_\mathrm{c}) = \langle
R_g^2 \rangle$.

G{\"u}rel and Gintoli investigated the relationship between
shear thinning and molecular conformation under shear flow.~\cite{gurel2023Shear}
Specifically, they examined the shear-rate dependence of viscosity
$\eta$ in 
polymers with linear, bottlebrush, and star-shaped architectures.
It was demonstrated that shear thinning behavior varies with 
chain architecture.
They further reported significant correlation 
coefficients between the shear viscosity $\eta$ of each polymer melt obtained and
several molecular conformation descriptors.
In particular, a strong correlation was observed between shear viscosity
$\eta$ and both the 
$y$-component of the gyration tensor, $\langle G_{yy}\rangle$, and the bond
orientational order parameter, $P_2$, defined by the second-order Legendre
polynomial.

Very recently, Uneyama analyzed the stress tensor and gyration tensor of
an unentangled polymer melt under flow using a Rouse-type single-chain
model.~\cite{uneyama2025Radius}
In this bead–spring model, the monomer beads are assumed to follow the
Langevin equation with a constant friction coefficient. 
He derived straightforward relations linking the stress tensor and the
gyration tensor in this Rouse-type model.
Under simple shear flow, the steady-state shear viscosity $\eta$ is governed by the
gyration radius along the velocity-gradient direction, specifically $\langle
G_{yy}\rangle$.

In this study, we extend previous investigation of the relationship between
shear viscosity $\eta$ and $\langle G_{yy}\rangle$ under steady shear flow
by exploring scaling relationships across different chain architectures
(linear and ring) and chain stiffness using NEMD simulations. 
We further analyzed the presence or absence of scaling relationships involving the
flow-direction component, $\langle G_{xx}\rangle$, and the component unaffected by advection, 
$\langle G_{zz}\rangle$, and examined their underlying mechanisms.
In addition, correlations between each component of the diagonalized
gyration tensor and shear viscosity were assessed.

\begin{figure}[t]
\centering
\includegraphics[width=0.5\textwidth]{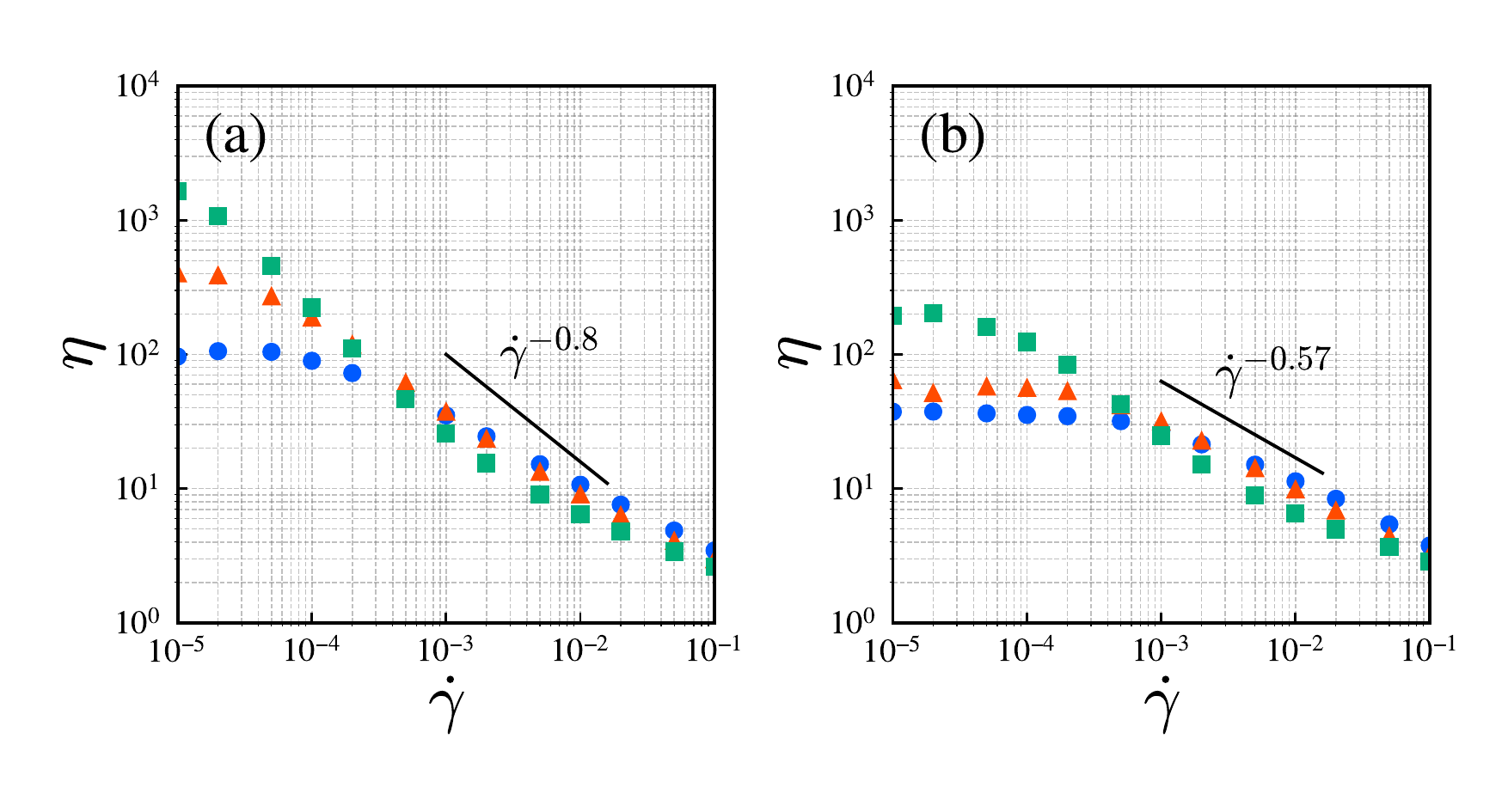}
\vspace*{-1cm}
\caption{Shear rate $\dot{\gamma}$ dependence of shear viscosity $\eta$
 for (a) linear (b) and ring polymers, with the chain stiffness
 $\varepsilon_\theta = 0$ (circle), 1.5 (triangle), and 3 (square).
The straight lines indicate the slopes corresponding to $0.8$ for
 linear polymers (a) and $0.57$
 for ring polymers (b), respectively.
}
\label{eta_gammadot}
\end{figure}

\begin{figure*}[t]
\centering
\includegraphics[width=0.8\textwidth]{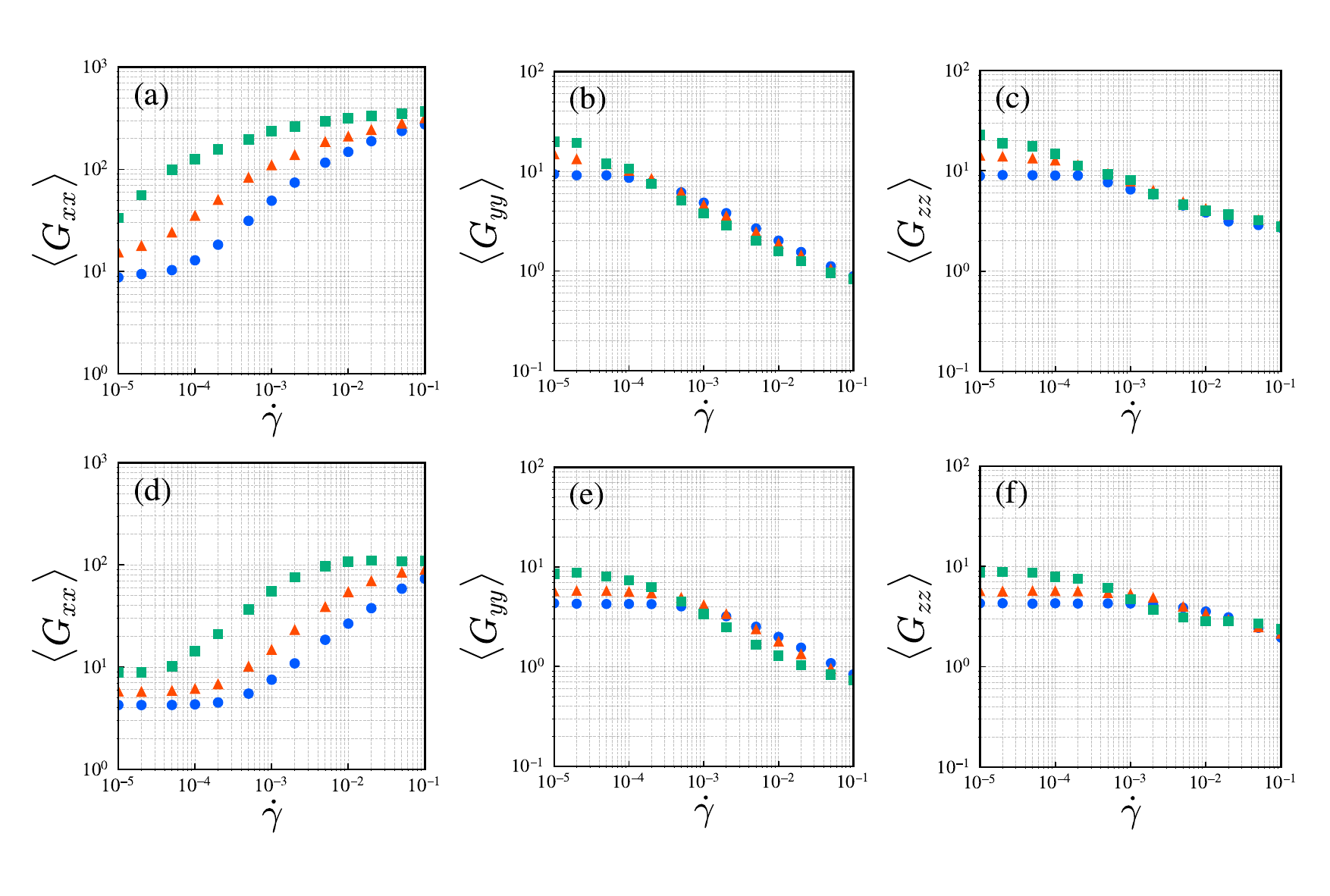}
\vspace*{-1cm}
\caption{Shear rate $\dot{\gamma}$ dependence of $\langle G_{xx}\rangle$ [(a) and (d)],
 $\langle G_{yy}\rangle$ [(b) and (e)], and $\langle G_{zz}\rangle$ [(c) and
 (f)] for
 linear [(a)-(c)] and ring [(d)-(f)] polymers.
Chain stiffness values are
 $\varepsilon_\theta = 0$ (circle), 1.5 (triangle), and 3 (square).
}
\label{G_gammadot}
\end{figure*}

\begin{figure*}[t]
\centering
\includegraphics[width=0.8\textwidth]{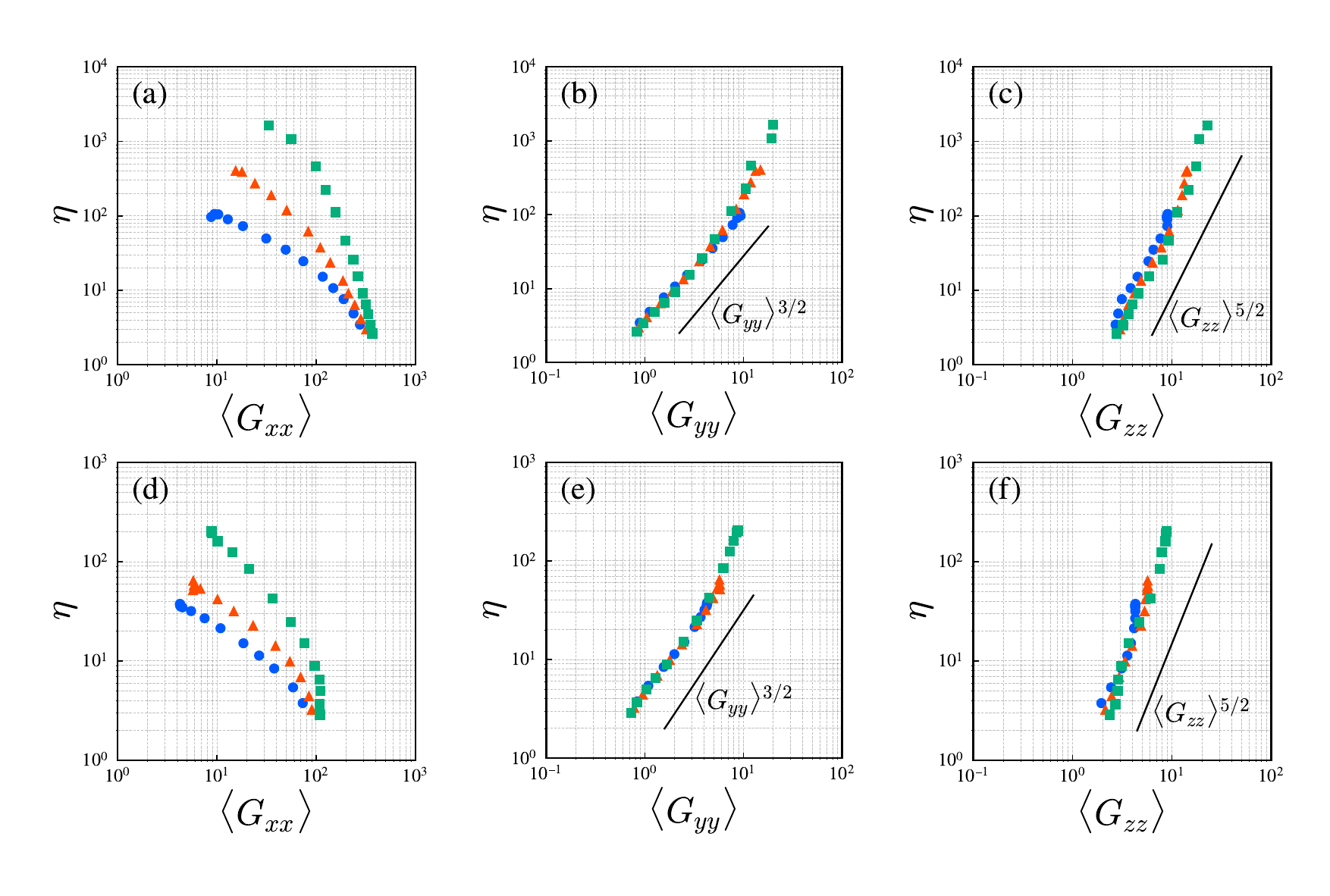}
\vspace*{-1cm}
\caption{Relationships between shear viscosity $\eta$ and $\langle G_{xx}\rangle$ [(a) and (d)],
 $\langle G_{yy}\rangle$ [(b) and (e)], and $\langle G_{zz}\rangle$ [(c) and
 (f)] for
 linear [(a)-(c)] and ring [(d)-(f)] polymers.
The straight lines indicate the slopes corresponding to $3/2$ in (b) and
 (e), and $5/2$ in (c) and (f), respectively.
Chain stiffness values are
 $\varepsilon_\theta = 0$ (circle), 1.5 (triangle), and 3 (square).
}
\label{G_eta}
\end{figure*}

\section{SIMULATION METHOD}

The Kremer-Grest model, a representative bead-spring model, was employed
to conduct NEMD simulations~\cite{kremer1990Dynamics}.
Each polymer chain is modeled by $N$ monomer beads of mass $m$ and
diameter $\sigma$. 
The system comprised $M$ polymer chains contained within a
three-dimensional cubic box with volume of $V$ and periodic boundary
conditions. 
Interactions between monomer beads were governed by three types of
interparticle potentials:
the Lennard-Jones (LJ) potential, applied to all bead pairs,
\begin{equation}
U_\mathrm{LJ}(r)  = 4\varepsilon_\mathrm{LJ}
 \left[\left(\frac{\sigma}{r}\right)^{12} -
  \left(\frac{\sigma}{r}\right)^{6}\right] + C,
\end{equation}
where $r$ denotes the distance between
two beads and $\varepsilon_\mathrm{LJ}$ specifies the LJ energy scale.
The LJ potential was truncated at a cut-off distance
$r_\mathrm{c}=2^{1/6} \sigma$, with the constant $C$ 
added to shift the potential such that $U_\mathrm{LJ}(r_\mathrm{c})=0$.
The second interparticle potential was the finite extensible nonlinear elastic
(FENE) model, applied between two adjacent monomer beads and expressed
as 
\begin{equation}
U_\mathrm{FENE}(r)  = -\frac{1}{2} K  R_0^2 \ln \left[1 - \left(\frac{r}{R_0}\right)^2\right],
\end{equation}
for $r< R_0$.
Here, $K$ represents the spring constant, and $R_0$ specifies the 
maximum bond extension between two beads.
The conventional parameter values
$K=30\varepsilon_\mathrm{LJ}/\sigma^2$ and $R_0=1.5\sigma$ were employed.
Finally, a bending elastic potential
\begin{equation}
U_\mathrm{bend}(\theta)=\varepsilon_\theta\left(1-\cos{\theta}\right)
\end{equation}
 was incorporated for the
bond angle $\theta$, 
with chain stiffness controlled by the parameter
$\varepsilon_\theta$.
It should be noted that the semiflexible chain with $\varepsilon_\theta/\varepsilon_\mathrm{LJ}=1.5$
is a commonly employed parameter in simulations of both
linear~\cite{hsu2016Static, hsu2017Detailed, goto2021Effects} and
ring~\cite{halverson2011Molecular, halverson2011Moleculara,
halverson2012Rheology, 
goto2021Effects, parisi2021Nonlinear} polymers.
Furthermore, the influence of the chain stiffness on the polymer
structure and dynamics was
investigated for both linear~\cite{faller1999Local, faller2000Local,
faller2001Chain, svaneborg2020Characteristica, everaers2020Kremer} and
ring~\cite{goto2023Unraveling, datta2023Viscosity, mei2024Unified} polymers.
Hereafter, length, energy, and time are expressed in units of
$\sigma$, $\varepsilon_\mathrm{LJ}$, and
$\sqrt{m\sigma^2/\varepsilon_\mathrm{LJ}}$, respectively, while 
temperature is given in units of $\varepsilon_\mathrm{LJ}/k_\mathrm{B}$.

In this study, 
NEMD simulations were performed by applying shear flow
to the melt of both linear and
ring molecular chains, with chain stiffness varied as 
$\varepsilon_\theta=0$, 1.5, and 3.
Note that the rheological entanglement length,
$N_\mathrm{e}^{\mathrm{rheo}}$, corresponding to the Kuhn length of primitive
paths, was estimated to be 80.5, 28.4, and 13.2 for
each value of $\varepsilon_\theta$ in the linear polymer systems, respectively.~\cite{dietz2022Validation}
The system was equilibrated at a fixed conditions of
temperature $T=1$
and a bead number density of $\rho=0.85$.
The number of beads per chain and the number of chains were set to $N=100$
and $M=100$, respectively. 
The equilibrium structure was used as the initial configuration, to
which 
a constant shear rate $\dot{\gamma}$ was applied. 
The SLLOD equations of motion were integrated under Lees-Edwards boundary conditions
to impose shear flow, with
the flow direction along the $x$-axis and the velocity-gradient
direction along
the $y$-axis.~\cite{evans2008Statistical}
The temperature was kept at $T=1$ using the Nos\'{e}--Hoover
thermostat during NEMD simulations.
The shear viscosity $\eta$ was calculated
from the $xy$ component of the stress tensor $\sigma_{xy}$ in the steady
state using the formula $\eta = \sigma_{xy}/\dot{\gamma}$.
The NEMD simulations in this study were carried out using 
the fix nvt/sllod command in
the Large-scale
Atomic/Molecular Massively Parallel Simulator (LAMMPS).~\cite{plimpton1995Fast}

\begin{figure*}[t]
\centering
\includegraphics[width=0.8\textwidth]{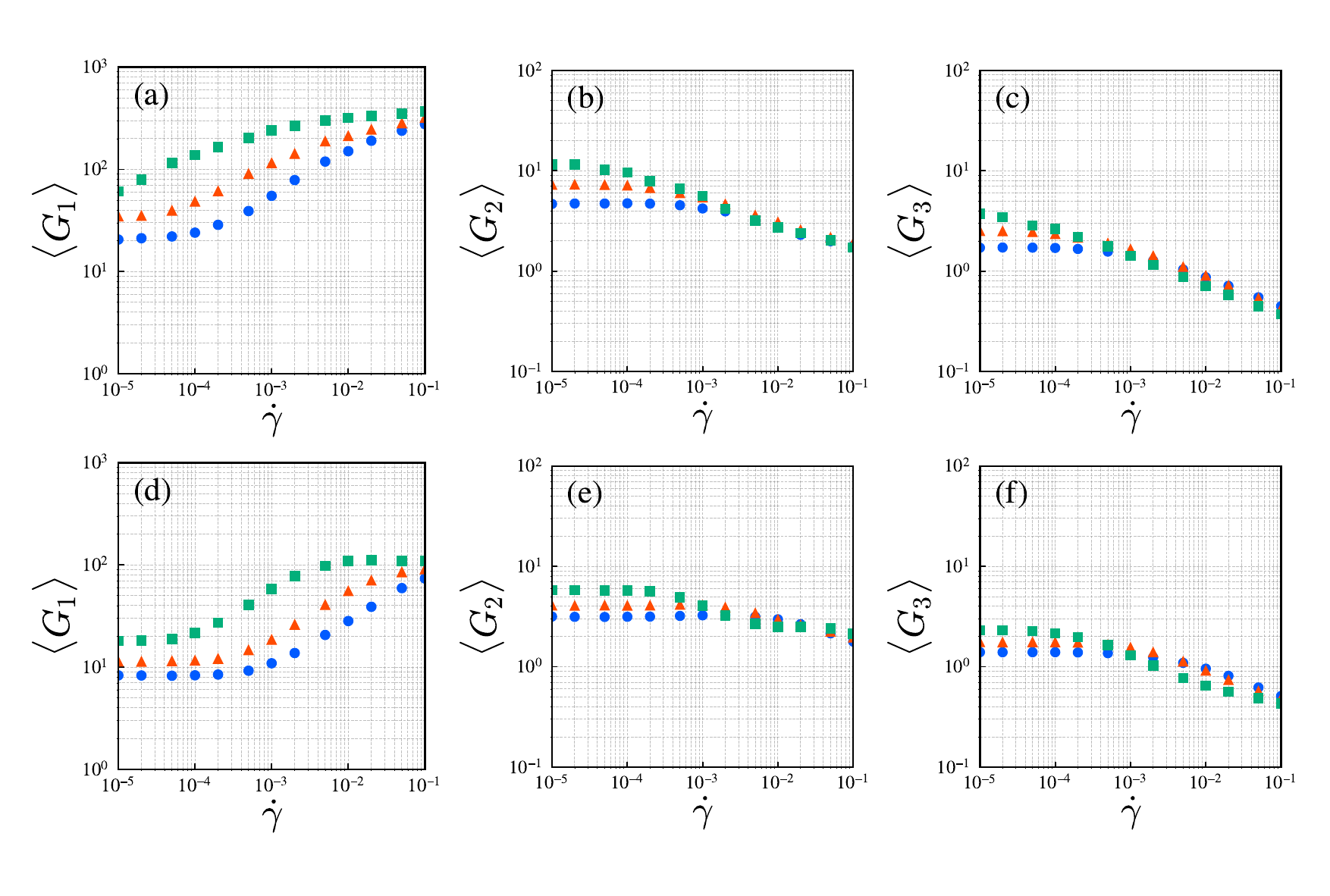}
\vspace*{-1cm}
\caption{Shear rate $\dot{\gamma}$ dependence of $\langle G_1\rangle$ [(a) and (d)],
 $\langle G_2\rangle$ [(b) and (e)], and $\langle G_3\rangle$ [(c) and
 (f)] for
 linear [(a)-(c)] and ring [(d)-(f)] polymers.
Chain stiffness values are
 $\varepsilon_\theta = 0$ (circle), 1.5 (triangle), and 3 (square).
}
\label{G1G2G3_gammadot}
\end{figure*}

\begin{figure*}[t]
\centering
\includegraphics[width=0.8\textwidth]{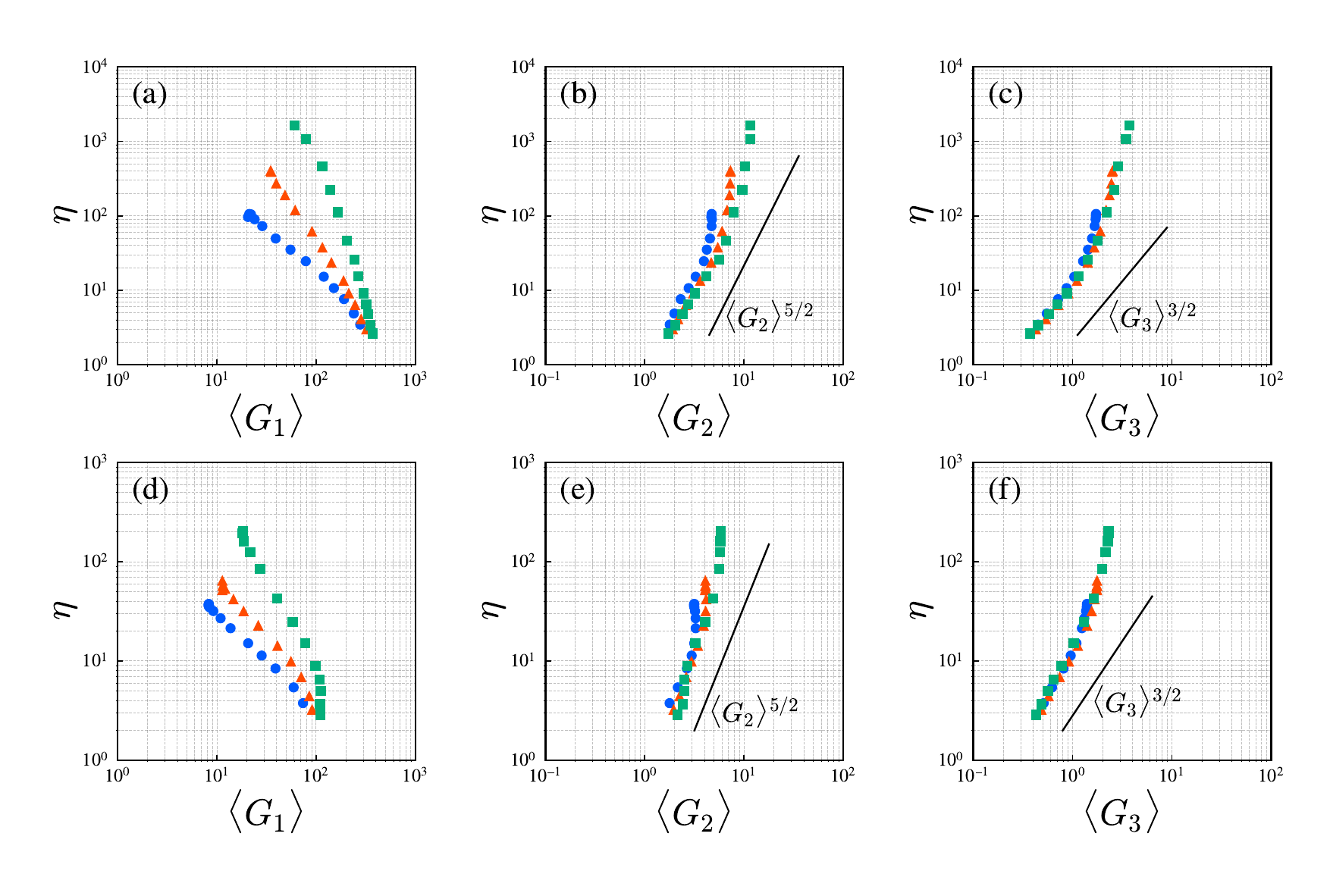}
\vspace*{-1cm}
\caption{Relationships between shear viscosity $\eta$ and $\langle G_1\rangle$ [(a) and (d)],
 $\langle G_2\rangle$ [(b) and (e)], and $\langle G_3\rangle$ [(c) and
 (f)] for
 linear [(a)-(c)] and ring [(d)-(f)] polymers.
The straight lines indicate the slopes corresponding to $3/2$ in (b) and
 (e), and $5/2$ in (c) and (f), respectively.
Chain stiffness values are
 $\varepsilon_\theta = 0$ (circle), 1.5 (triangle), and 3 (square).
}
\label{G1G2G3_eta}
\end{figure*}

\section{RESULTS AND DISCUSSION}

First, we show the results of shear-thinning behavior in linear and ring
polymers by varying the chain stiffness $\varepsilon_\theta$.
Figure~\ref{eta_gammadot} presents the simulation results of the shear
viscosity $\eta$ as a function of shear rate $\dot{\gamma}$ in both
linear and ring polymers.
Parisi \textit{et al.} reported NEMD simulation results for $\eta$ as a
function of 
$\dot{\gamma}$ for linear and ring polymers with chain stiffness
$\varepsilon_\theta = 1.5$, varying the chain length $N$.~\cite{parisi2021Nonlinear}
The results for $\varepsilon_\theta = 1.5$ with $N=100$, shown in
Fig.~\ref{eta_gammadot}, are consistent with those reported in their study.
When comparing linear and ring polymers, it was observed that ring
polymers exhibit lower viscosity than their linear counterparts at low
shear rates.
This behavior can be attributed to the absence of chain ends in ring
polymers and the enhanced chain tumbling dynamics, which lead to lower
viscosity compared to linear polymers.~\cite{parisi2021Nonlinear}
Notably, the tumbling events under shear flow were directly
identified by analyzing the time correlation function of the end-to-end
vector for a chain section.
In particular, the sudden decay of time correlation along the flow
direction 
was found to characterize
the tumbling motions.~\cite{parisi2021Nonlinear}
In contrast, 
the values of $\eta$ for linear and ring polymers converge as the shear rate
increases. 
This trend is also consistent with the previous study.~\cite{parisi2021Nonlinear}
Furthermore, the shear-thinning behavior was characterized by 
exponent of $\alpha=0.80$ for linear polymers and $\alpha=0.57$ for ring
polymers.~\cite{parisi2021Nonlinear}
Figure~\ref{eta_gammadot} also demonstrates the shear-thinning behavior in
the large-$\dot{\gamma}$ regime, with consistent exponent values
observed for linear
and ring polymers.
Focusing on the dependence on $\varepsilon_\theta$ in
Fig.~\ref{eta_gammadot}, both linear and polymer chains
exhibit an increase in $\eta$ and enhanced non-linearity as
$\varepsilon_\theta$ increases to 3, leading to the onset of shear thinning at 
lower shear rate regimes.

We calculated the gyration tensor for each polymer chain, defined as
\begin{equation}
G_{\alpha\beta} = \frac{1}{N} \sum_{i=1}^{N}
 (\alpha_i-\alpha_\mathrm{CM}) (\beta_i-\beta_\mathrm{CM}),
\end{equation}
where $\alpha_i$ ($\beta_i$) denotes the $\alpha$ ($\beta$) component of monomer bead $i$ with
$\alpha, \beta= (x, y, z)$.
Furthermore, $\alpha_\mathrm{CM}$ ($\beta_\mathrm{CM}$) represents
$\alpha$ ($\beta$) component of the polymer chain's center of mass.
The square of gyration radius $R_g^2$ is calculated as the sum of
the eigenvalues of the gyration tensor: $R_g^2=G_1+G_2+G_3$, where the
principal axes are chosen 
such that the diagonal elements are ordered as 
$G_1 \ge G_2 \ge G_3$.

Figure~\ref{G_gammadot} represents the shear rate $\dot{\gamma}$
dependence of $\langle G_{xx} \rangle$, $\langle G_{yy} \rangle$, and
$\langle G_{zz} \rangle$ for linear and ring polymers.
Note that $\langle \cdots \rangle$ denotes the average over chains and time.
For both linear and ring polymers, increasing the shear rate
$\dot{\gamma}$ leads to an increase in 
$\langle G_{xx}\rangle$, which eventually converges to a constant value
independent of $\varepsilon_\theta$.
This behavior arises because monomer beads are advected along the flow
direction ($x$-axis), causing polymer chains to elongate. 
Furthermore, the $\langle G_{xx}\rangle$ value 
for ring polymers is smaller than that for linear polymers at any given
$\dot{\gamma}$ and chain stiffness $\varepsilon_\theta$, 
which can be attributed to the chain bending inherently
induced by the absence of chain ends.
In contrast, for both linear and ring polymers, $\langle G_{yy}\rangle$
decreases with increasing $\dot{\gamma}$.
This behavior is attributed to the advection of monomer beads along the
flow direction, which limits chain extension along the
velocity-gradient direction ($y$-axis).
Comparing linear and ring polymers, $\langle G_{yy}\rangle$ is 
smaller for ring polymers, particularly at low shear rates.
Similar to $\langle G_{xx}\rangle$, this indicates ring polymers extend
less readily than linear polymers due to the absence of chain ends.
$\langle G_{zz}\rangle$ decreases with increasing $\dot{\gamma}$,
exhibiting a behavior similar to that of $\langle G_{yy}\rangle$ for
both linear and ring polymers.
However, $\langle G_{zz}\rangle$ remains larger than $\langle
G_{yy}\rangle$ at any given $\dot{\gamma}$ and chain stiffness
$\varepsilon_\theta$.
This result can be explained by the fact that polymer chains can extend
more freely in the $z$-direction, where advection effects are absent.

We examine the relationships between shear viscosity $\eta$ and each
component of average gyration tensor, namely, $\langle G_{xx}\rangle$,
$\langle G_{yy}\rangle$, and
$\langle G_{zz}\rangle$, in Fig.~\ref{G_eta}.
Notably, the scaling relation $\eta \propto \langle
G_{yy}\rangle^{3/2}$ is observed in Fig.~\ref{G_eta}(b) and (e),
independent of chain architecture and
chain stiffness, particularly in the $\eta \ls 100$ regime for linear
polymers and $\eta \ls 50$ regime for ring polymers.
This scaling behavior is consistent with previously reported
results,~\cite{xu2017Probing} and quantitatively indicates that 
the extension of polymer chain in the velocity-gradient
direction, as characterized by $\langle G_{yy} \rangle$.
Assuming that each chain is dynamically confined within
a distance of order $\langle G_{yy}\rangle $ along the
velocity-gradient direction $y$, the
drag force in this direction can be directly associated with the shear
viscosity $\eta$.
In contrast, the scaling exponent becomes
greater than $3/2$ in the higher-viscosity regime, which
corresponds to low shear rates, suggesting the entanglement
effects contribute to the increase in $\eta$.
Additionally, Fig.~\ref{G_eta}(c) and (f) shows a similar scaling relation,
$\eta \propto \langle G_{zz}\rangle^{5/2}$, which is independent of chain
architecture and chain stiffness in the $\eta \ls 100$ regime for linear
polymers and $\eta \ls 50$ regime for ring polymers.
Note that the scaling exponent exceeds $5/2$ in the higher-viscosity regime,
similar to the behavior observed for the scaling with $\langle G_{yy}\rangle$.
The underlying basis of the observed scaling relation can be understood
as follows. 
As mentioned above, there is no advection in the $z$-direction, meaning
that each polymer chain can be considered to diffuse freely along this direction.
In other words, the diffusion of polymer chains in this
direction is expected to remain nearly identical to that in equilibrium.
Consequently, the magnitude of $\langle G_{zz}\rangle$ characterizes the
hydrodynamic size of the chains, and 
determines their diffusion coefficient along the
$z$-direction, which is also related to $\eta$ under the assumption of the
Stokes--Einstein relation.
For reference, in the case of a linear Rouse chain at equilibrium, 
$\langle R_g^2\rangle$ is proportional to the chain length $N$ and
inversely proportional to the diffusion coefficient, $D$. 
Furthermore, by applying the Stokes–Einstein relation, it follows 
$\eta \propto \langle R_g^2\rangle$, consistent with the
observation of Xu \textit{et al}.~\cite{xu2015Simulation}
However, the deviation from the power of $1$ observed in this study is
likely due to the breakdown of both the Rouse chain assumption and
the Stokes–Einstein relation.
Finally, it was found that no clear scaling relationship exists between
$\eta$ and $\langle G_{xx} \rangle$ in Fig.~\ref{G_eta}(a) and (d).
This can be explained by Fig.~\ref{eta_gammadot}(a) and (b) and
Fig.~\ref{G_gammadot}(a) and (c), which show that at low shear rates,
$\eta$ and $\langle G_{xx}\rangle$ differ due to chain stiffness
$\varepsilon_\theta$, whereas, as indicated in Fig.~\ref{G_gammadot}, $\langle G_{xx}\rangle$
converges to a constant value at higher shear rates for both linear and ring polymers.

Thus far, we have presented the results for the gyration tensor components in the
laboratory coordinate system.
Finally, we show the results obtained from analyzing the diagonalized
gyration tensor components, $G_1$, $G_2$, and $G_3$, corresponding to
the squared lengths of 
the semi-axes when the polymer chain is regarded as an ellipsoid.
Figure~\ref{G1G2G3_gammadot} presents the dependence of $\langle G_1\rangle$, $\langle G_2\rangle$, and $\langle
G_3\rangle$ on the shear rate $\dot{\gamma}$
for both linear and ring polymers.
The results indicate that
$\langle G_1\rangle$, $\langle G_2\rangle$, and $\langle G_3\rangle$
correspond respectively to 
$\langle G_{xx}\rangle$, $\langle G_{zz}\rangle$, and  $\langle
G_{yy}\rangle$.
This implies that, for both linear and ring polymers under shear flow,
the long axis aligns with the flow direction along the $x$-axis.
Furthermore, as shown in Fig.~\ref{G_gammadot}, the
chains tend to extend more readily along the $z$-axis, which is less
affected by advection than the velocity-gradient direction along the
$y$-axis, accounting for the correspondence between $\langle G_2 \rangle$ and
$\langle G_{zz}\rangle$.
The relationship with the shear viscosity $\eta$ is illustrated in Fig~\ref{G1G2G3_eta}.
This results suggests that similar scaling relationships apply to the
diagonalized gyration tensor components, $\langle G_2\rangle$ and
$\langle G_3\rangle$, namely, $\eta \propto \langle G_2 \rangle^{5/2}$, and
$\eta \propto \langle G_3\rangle^{3/2}$.
However, in the high-viscosity regime corresponding to low shear rates,
deviations from the scaling
behaviors are observed. 
This can be attributed to fluctuations that prevent the long axis from
fully aligning with the flow direction along the $x$-axis.

\section{CONCLUSIONS}

In this study, we investigated the correlation between shear thinning and
the gyration tensor in polymer melts with varying chain stiffness
$\varepsilon_\theta$ for
both linear and ring polymers, using NEMD
simulations based on the Kremer–Grest model. 
We observed the previously
reported scaling relation, $\eta\propto \langle G_{yy}\rangle^{3/2}$, as
well as an additional scaling relation, $\eta\propto \langle G_{zz}\rangle^{5/2}$.
Furthermore, similar scaling relationships were found for the diagonalized gyration tensor components,
$\langle G_2\rangle$ and $\langle G_3\rangle$.
These results indicate that the degree of molecular extension, particularly along the
velocity-gradient direction and the direction unaffected by advection,
accounts for the structural influence on shear thinning in both linear and
ring polymers.
To clarify the observed scaling relationship
between $\eta$ and $\langle G_{zz}\rangle$, it is important
to analyze the shear rate dependence of the diffusion coefficient of
polymer chains under shear flow, which becomes anisotropic.
In particular, by examining the relationship between the $z$-component, which is
unaffected by the advection, and $\eta$, the validity of the
Stokes--Einstein relation can be assessed.
Our research is currently progressing in this direction.
Furthermore, analyses including bottlebrush polymers have shown that the
orientation of individual bonds, specifically quantified by $P_2$, correlates more strongly with
architecture-dependent shear-thinning behavior than with the components
of the gyration tensor, both before and after diagonalization.~\cite{gurel2023Shear}
Further investigation into the factors governing shear-thinning behavior
with varying polymer architectures is important and should be addressed
in future studies.

\begin{acknowledgments}
This work was supported by 
JSPS KAKENHI Grant-in-Aid 
Grant Nos.~\mbox{JP25K00968}, \mbox{JP24H01719}, \mbox{JP22K03550}, and \mbox{JP23H02622}.
We acknowledge support from
the Fugaku Supercomputing Project (Nos.~\mbox{JPMXP1020230325} and \mbox{JPMXP1020230327}) and 
the Data-Driven Material Research Project (No.~\mbox{JPMXP1122714694})
from the
Ministry of Education, Culture, Sports, Science, and Technology.
The numerical calculations were performed at Research Center of
Computational Science, Okazaki Research Facilities, National Institutes
of Natural Sciences (Project: 25-IMS-C052).
\end{acknowledgments}

%aipnum4-2.bst 2019-01-14 (MD) hand-edited version of apsrev4-1.bst

\end{document}